\documentclass{emulateapj}
\usepackage{color}

\definecolor{dr}{rgb}{0.6,0,0}
\definecolor{db}{rgb}{0,0,0.6}

\usepackage[ hyperfootnotes={false},
dvips,
pdftitle={Planet formation around stars of various masses: Hot super-Earths},
pdfauthor={Grant Kennedy},
pdfstartview={FitH},
bookmarksopen={true} ]{hyperref}

\hypersetup{colorlinks={true}, urlcolor={db}, citecolor={dr}}

\begin{document}

\submitted{Accepted to ApJ}

\title{Planet formation around stars of various masses: Hot super-Earths}

\author{Grant M. Kennedy}%
\affil{Research School of Astronomy and Astrophysics, Mt Stromlo
  Observatory, Australian National University, ACT 2611, Australia}
\email{grant@mso.anu.edu.au}%

\and

\author{Scott J. Kenyon}
\affil{Smithsonian Astrophysical Observatory, Cambridge, MA 02138, USA}
\email{kenyon@cfa.harvard.edu}

\shortauthors{KENNEDY \& KENYON}
\shorttitle{HOT SUPER-EARTHS}

\begin{abstract}
  We consider trends resulting from two formation mechanisms for
  short-period super-Earths: planet-planet scattering and
  migration. We model scenarios where these planets originate near the
  snow line in ``cold finger'' circumstellar disks. Low-mass
  planet-planet scattering excites planets to low periastron orbits
  only for lower mass stars. With long circularisation times, these
  planets reside on long-period eccentric orbits. Closer formation
  regions mean planets that reach short-period orbits by migration are
  most common around low-mass stars. Above $\sim$1\,$M_\odot$, planets
  massive enough to migrate to close-in orbits before the gas disk
  dissipates are above the critical mass for gas giant
  formation. Thus, there is an upper stellar mass limit for
  short-period super-Earths that form by migration. If disk masses are
  distributed as a power law, planet frequency increases with
  metallicity because most disks have low masses. For disk masses
  distributed around a relatively high mass, planet frequency
  decreases with increasing metallicity. As icy planets migrate, they
  shepherd interior objects toward the star, which grow to
  $\sim$1\,$M_\oplus$. In contrast to icy migrators, surviving
  shepherded planets are rocky. Upon reaching short-period orbits,
  planets are subject to evaporation processes. The closest planets
  may be reduced to rocky or icy cores. Low-mass stars have lower EUV
  luminosities, so the level of evaporation decreases with decreasing
  stellar mass.
\end{abstract}


\section{Introduction}\label{sec:intro}

With nearly 300 known extra-Solar planets, there are now several clear
correlations between the properties of the planets and their host
stars. The most well known trend is the increase in gas giant
frequency with host star metallicity
\citep[e.g.][]{2005ApJ...622.1102F}. Recent radial velocity surveys
suggest that giant planet frequency also increases with stellar mass
\citep{2007ApJ...670..833J}.

These trends provide tests of planet formation theories. In the core
accretion model for example, gas giant planets form by coagulation of
small planetesimals near the ``snow line'' that separates rocky and
icy regions in a circumstellar disk. Once icy protoplanets reach a
critical core mass, they accrete gas rapidly
\citep{1996Icar..124...62P}. Cores benefit from extra planet building
material provided by enhanced metallicities and an increase in disk
masses with stellar mass. The model is thus consistent with current
observations
\citep{2004ApJ...616..567I,2005ApJ...626.1045I,2008ApJ...673..502K}.

Gravitational instability (GI) is an alternative formation mechanism
for gas giant planets, where a relatively massive disk cools enough to
fragment into Jupiter-mass clumps. Although GI operates over a wide
range of stellar masses, there is still debate about predicted trends
with metallicity \citep{2007prpl.conf..607D}. Given observational
biases in the current sample of extra-Solar planets, GI cannot be
ruled out as a formation mechanism \citep{2007prpl.conf..607D}.

Core accretion and GI models suggest that short-period ``hot
Jupiters'' reside too close to their parent stars to have formed
\emph{in situ}. Thus, these planets must migrate or scatter from more
distant formation regions to arrive at their final orbits
\citep{1996Natur.380..606L,1996Sci...274..954R}. A combination of
these two mechanisms probably operates to produce the observed
distribution of extra-Solar giant planets. Scattering can reproduce
most of the observed eccentricity distribution, but has trouble
accounting for planets in circular orbits at distances too far from
their host stars for tidal circularisation
\citep{2007astro.ph..3163F}. Migration theories can explain systems
with planets in mean-motion resonances \citep{2002ApJ...567..596L},
but they may not reproduce the observed eccentricity distribution
\citep[e.g.][]{2004AIPC..713..243T}.

With the discovery of the first super-Earths in relatively short
period orbits, migration and scattering remain possible mechanisms for
planets to reach these radii
\citep{2005Icar..177..264B,2007ApJ...654.1110T,2007arXiv0711.2015R}. However,
the discovery of low-mass planets in systems already harbouring giant
planets suggests new formation mechanisms
\citep{2005ApJ...631L..85Z}. Because these models require gas giants,
they predict trends with metallicity and stellar mass for low-mass
planets similar to those for giant planets. Though some low-mass
planets may have formed with help from giant planets, a flatter
metallicity distribution \citep{2007prpl.conf..685U} and the absence
of giant planets in some low-mass planet systems \citep[e.g. Gl581 \&
GJ674,][]{2005A&A...443L..15B,2007A&A...474..293B} indicate other
formation mechanisms.

Here, we consider trends that may arise in forming short-period and/or
transiting icy/rocky planets in systems with no gas giants, over a
range of stellar masses. The close-in planets that form are therefore
the most massive in the planetary system. We first cover some
background in \S \ref{sec:background}. In \S \ref{sec:scattering} we
use $n$-body simulations to show that 10\,$M_\oplus$ planet-planet
scattering is unlikely to result in transiting planets for all but the
lowest mass stars. With long circularisation timescales, planets in
these systems are hard to detect. We consider migration scenarios
using analytic, semi-analytic and $n$-body models in \S
\ref{sec:migration}. With migration, short-period low-mass planets
most likely form around low-mass stars. Above a certain stellar mass,
it is hard to form any short-period planets without giant
atmospheres. Trends with metallicity depend on the disk mass
distribution. Migration to short-period orbits results in significant
amounts of material being shepherded inward, which affects the final
structure of these systems. We discuss our results, subsequent
planetary evolution, and conclude in \S \ref{sec:summary}.

\section{Background}\label{sec:background}

\subsection{General Picture}\label{sec:gen-background}

Planets form in circumstellar disks. Therefore disk structure plays a
key role in setting the final configuration of planetary systems. In
most planet formation models, disk structure is characterised by an
outwardly decreasing radial surface density profile. This profile
usually includes an increase in surface density at the ``snow line,''
where the temperature becomes low enough for water to freeze.

Planets form by accumulating solids in the disk. Therefore the
expected increase in surface density at the snow line is often
associated with the formation of gas giants like Jupiter. Forming
Jupiter requires the relatively rapid growth of a
$\sim$5--10\,$M_\oplus$ icy core, followed by a period of gas
accretion \citep{1996Icar..124...62P}. Gas accretion must be complete
before the gas disk disperses in $\sim$3\,Myr
\citep[e.g.][]{2001ApJ...553L.153H}. In the minimum mass Solar nebula
model \citep[MMSN,][]{1977Ap&SS..51..153W,1981PThPS..70...35H},
forming the icy core rapidly requires factor of 5--10 surface density
enhancements relative to the terrestrial region
\citep{1987Icar...69..249L,1996Icar..124...62P,2003Icar..161..431T}. This
factor is larger than the factor of 2--3 enhancements expected from
Solar abundances \citep{2005ASPC..336...25A}, or suggested by comet
composition \citep{2005Natur.437..987K}, and the factor of $\sim$4
derived in the original MMSN model \citep{1981PThPS..70...35H}.

The need for larger surface density enhancements inspired ``cold
finger'' disk models, which produce much larger snow line enhancements
in a relatively narrow ($\lesssim$AU) radial region near the snow line
\citep{1988Icar...75..146S,2004ApJ...614..490C}. In this picture, a
circumstellar disk has an initial equilibrium state with the water
vapour (ice) concentration decreasing (increasing) beyond the snow
line. As the disk diffuses and advects, water continually condenses
from gas passing beyond the snow line, thus enhancing the local
surface density of solids, and removing vapour phase water from the
inner disk. Sublimation of planetesimals that drift inside the
condensation radius by gas drag enhances this effect: the surface
density beyond the snow line increases when water vapour from the
sublimated planetesimals diffuses back outside the snow line
\citep{2004ApJ...614..490C}.

The first cold finger models predict a factor of $\sim$10-100 increase
in the surface density of icy material in a relatively narrow region
near the snow line
\citep{1988Icar...75..146S,2004ApJ...614..490C}. Using a more complex
global disk model, \citet{2006Icar..181..178C} suggest surface density
enhancements closer to 10 than 100. In their simulations, the
enhancement regions are several AU wide at half the maximum
planetesimal surface density.

The main differences expected for planet formation models using cold
finger instead of MMSN disks are threefold. Due to the nature of the
surface density enhancement: (i) fewer large planets form, (ii) large
planets form in relatively low-mass disks, and (iii) planets form from
material with much higher ice/rock ratios. In addition, material lost
to inward planetesimal drift by gas drag \citep{2003Icar..161..431T}
may be returned to the cold finger region, allowing continued
growth. Reducing the removal of drifting planetesimals enhances growth
rates, and allows formation of more massive icy planets.

\subsection{Mathematical Formalism}\label{sec:math-background}

In the standard coagulation model, planets grow in a circumstellar
disk through repeated collisions and mergers of smaller objects
\citep{1969QB981.S26......}. First, roughly km size planetesimals
form rapidly, whether by coagulation
\citep[e.g.][]{2000SSRv...92..295W} or direct collapse
\citep[e.g.][]{1973ApJ...183.1051G}. Little knowledge of which process
dominates means the size distribution of the first planetesimals is
poorly constrained. Planetesimals initially grow through a rapid phase
of ``runaway'' growth \citep{1996Icar..123..180K}. During the period
of ``oligarchic'' growth that follows \citep{1998Icar..131..171K},
protoplanetary growth rates depend on the surface density of
planetesimals $\sigma_{\rm s}$, the local orbital frequency $\Omega$,
the gravitational reach of the growing protoplanet, and the random
velocities of the smaller planetesimals \citep{2001Icar..149..235I}
\begin{equation}\label{eq:mdot}
  \dot{M_{\rm pl}} \propto \sigma_{\rm s} \, r_{\rm H}^2 \, \Omega
  \, P_{\rm col}(\tilde{e},\tilde{i}) \, .
\end{equation}
Here $r_{\rm H} = a \left( M_{\rm pl} / 3 M_\star \right)^{1/3}$ is
the Hill radius, and $a$ is semi-major axis. The eccentricity
$\tilde{e}$ and inclination $\tilde{i}$ are in units of the growing
protoplanets Hill radius (i.e. $\tilde{e} = e/r_{\rm H}$). The
collision probability $P_{\rm col}$ largely determines how growth
proceeds: growth is fastest when planetesimals are small enough
($\lesssim$1\,km) to be damped by gas drag
\citep[e.g.][]{2004AJ....128.1348R}. In this ``shear dominated''
regime when $\tilde{e}$ and $\tilde{i}$ are $\lesssim$1, growth
depends on Keplerian shear in the disk, rather than objects random
velocities. Growth slows strongly with increasing radial distance,
because $\Omega \propto a^{-3/2}$ and $\sigma_{\rm s} \propto
a^{-\delta}$, where $\delta \sim 1$--1.5.

Eventually, protoplanets accrete most of the nearby material and reach the
``isolation'' mass \citep{1987Icar...69..249L}
\begin{equation}\label{eq:miso1}
  M_{\rm iso} = \frac{ \left( 4 \pi B \sigma_{\rm s} a^2 \right)^{3/2} }
  { \left( 3 M_\star \right)^{1/2} } \, .
\end{equation}
Numerical simulations indicate that isolated oligarchs are spaced at
$2B R_{\rm H} \sim 8 R_{\rm H}$ intervals
\citep[e.g.][]{1998Icar..131..171K}. In the terrestrial region around
the Sun, the isolation mass is $\sim$0.1\,$M_\oplus$, and the
timescale for Earth formation by the chaotic growth that follows is
$\sim$10-100\,Myr \citep[e.g.][]{2006AJ....131.1837K}.

Further out in the disk, larger isolation masses allow formation of
gas giant planets. The critical core mass for gas accretion depends on
opacity and planetesimal accretion rates, but is
$\gtrsim$10\,$M_\oplus$
\citep[e.g.][]{2000ApJ...537.1013I,2006ApJ...648..666R}. This mass is
reached more easily further out in the disk because $M_{\rm iso}$
increases with $a$. However, growth slows rapidly with increasing
radial distance; thus, there is an optimum region where cores are
massive enough to accrete gas and to form giant planets before the gas
disk is dissipated \citep{2008ApJ...673..502K}. This region is
sufficiently far from the star that \emph{in situ} formation of
``hot-Jupiters'' is unlikely, thus motivating theories of migration
and scattering.

\subsection{Migration}\label{sec:mig-background}

Type I migration is a potential barrier to the formation of both
terrestrial and giant planets
\citep{1980ApJ...241..425G,1997Icar..126..261W,2002ApJ...565.1257T,2007prpl.conf..655P}.
When protoplanets reach near an Earth mass, the excitation of spiral
density waves in the gaseous disk causes planets to experience a
torque, and migrate inward. The timescale for a planet to spiral into
the central star is \citep{2002ApJ...565.1257T}
\begin{equation}\label{eq:taumig}
  \tau_{\rm mig} = \left( 2.7 + 1.1 \delta \right)^{-1}
  \frac{ \left( M_\star M_\odot \right)^2 }{ M_{\rm pl} \, \sigma_{\rm
      gas} \, a^2 } \frac{ h^2 }{ \Omega } \, ,
\end{equation}
where $h \approx 0.05$ is the disk aspect ratio, and the stellar mass
$M_\star$ is in units of Solar masses. For a planet of mass $M_{\rm
  pl} = 1\,M_\oplus$ in a disk with $\sigma_{\rm gas} = 1700$\,g
cm$^{-2}$ at 1\,AU around a Solar-mass star, $\tau_{\rm mig} = 1.6
\times 10^5$\,yr. Because this timescale is shorter than the
$\sim$3\,Myr disk lifetime \citep{2001ApJ...553L.153H}, and comparable
with growth timescales, type I migration theory conflicts with
terrestrial and giant planet formation in the Solar System \citep[but
see][]{2006ApJ...652L.133C}.

Recent work suggests a reduced migration efficiency can resolve this
problem \citep{2008ApJ...673..487I}. This ``offset'' applies to
planets $\lesssim$15\,$M_\oplus$
\citep{2002A&A...385..647D,2003ApJ...586..540D} and arises from
corotation torques by coorbital material
\citep{2006ApJ...652..730M}. Other ways of reducing (and even
reversing) type I migration rates include turbulence arising from the
magnetorotational instability \citep[e.g.][]{2004MNRAS.350..849N}, and
eccentricity driven by planet-planet interactions
\citep{2000MNRAS.315..823P}.

If planets do not fall onto the central star, migration is a possible
mechanism for producing planets on short-period orbits
\citep{1996Natur.380..606L,2005Icar..177..264B,2007ApJ...654.1110T}.

\subsection{Scattering}\label{sec:scattering-background}

Planet-planet scattering can also produce planets with short-period,
or low periastron ($q$) orbits. Originally proposed to explain
hot-Jupiters \citep{1996Sci...274..954R}, this scenario has not been
applied to low-mass planets.

Scattering favours giant planets on short-period orbits. When a gas
giant scatters into a low periastron orbit, tidal interaction with the
star can circularise the orbit on reasonable timescales, with $a \sim
2q$ \citep{1996Sci...274..954R}. For lower mass planets, long
circularisation timescales make circular orbits unlikely
\citep{2007arXiv0711.2015R}. However, if the initial scattering region
is sufficiently close, as for low-mass stars, detection of
low-periastron eccentric planets is possible.

We now consider two different scenarios that form short-period and/or
transiting low-mass planets that begin growth near the snow line,
across a range of stellar masses. When the snow line enhancement is
small, many planets migrate toward close orbits. This scenario has
already been studied for Solar-mass stars by
\citet{2007ApJ...654.1110T}. Here, we instead consider cold finger
type disks, where a few planets forming near the snow line dominate
others forming elsewhere in the disk. We first consider a scattering
scenario resulting from \emph{in situ} growth, and then a migration
scenario. We defer discussion of subsequent planetary evolution in
final orbits to \S \ref{sec:summary}.

\section{Scattering}\label{sec:scattering}

Planet-planet scattering is a likely outcome of oligarchic growth. In
migration scenarios, protoplanets interact strongly with the gas disk,
and they migrate to close-in orbits. However, if the gas disk
disperses before planets have time to migrate, or if migration results
in no net inward movement, planets form \emph{in situ}. During
oligarchic growth, protoplanets grow on orbits near the limits of
dynamical stability, with damping provided by small bodies
\citep[e.g.][]{1988Icar...74..542S,1998Icar..131..171K}. At later
stages near isolation, their orbits can become unstable as remaining
small bodies are accreted
\citep{2004ApJ...614..497G,2006AJ....131.1837K}.

When planets start interacting dynamically, the boundary in semi-major
axis between stable and unstable configurations is very sharp. Thus,
two planets with orbits that become too close experience the sudden
onset of a dynamical instability caused by close encounters
\citep{1993Icar..106..247G}.

In previous studies of giant planet scattering, planets begin at
$\sim$AU distances from the central star, with spacings just inside
the stability limit. After many interactions, one planet sometimes
attains a highly eccentric orbit with a small periastron distance
\citep[e.g.][]{1996Sci...274..954R,2007astro.ph..3163F}. Tidal
interaction with the central star then circularises the orbit with $a
\sim 2q$.

While tidal forces can circularise gas giant orbits, the timescales
for 1--10\,$M_\oplus$ planets on highly eccentric orbits are long
\citep[$\gtrsim$Gyr,][]{2007arXiv0711.2015R}. Although these planets
maintain eccentric long-period orbits, transits are possible in
favourable circumstances. Because planets form at shorter orbital
periods around low-mass stars, these provide the best opportunity for
transit observations.

Cold finger disks provide an ideal environment for oligarchic growth
followed by planet-planet scattering. The width of the cold finger
region allows several protoplanets to form
\citep{2006Icar..181..178C}. Once protoplanets reach isolation,
further chaotic growth may occur if their escape velocity $v_{\rm
  esc}$ is less than the local Keplerian velocity $v_{\rm K}$
\citep[$\mathcal{R} \equiv v_{\rm esc}/v_{\rm
  K}$,][]{2004ApJ...614..497G}. In the terrestrial region of
Solar-type stars, $\mathcal{R} \sim 1/4$. For gas giants, $\mathcal{R}
\gg 1$. For $M_{\rm pl} = 10\,M_\oplus$ with density $\rho = 4.5$\,g
cm$^{-2}$, $\mathcal{R} \approx 1.3$ outside the snow line. Thus,
$\sim$10\,$M_\oplus$ protoplanets present an approximate division
between coalescence and scattering/ejection, and an order of magnitude
estimate of the maximum planet mass. This mass is similar to the
minimum needed for gas accretion, so scattering of super-Earths to
close-in orbits appears difficult.

For less massive stars, scattering to low periastron orbits is
easier. At fixed $a$, smaller $v_{\rm K}$ leads to larger
$\mathcal{R}$ and a greater chance of scattering. However, the snow
line also moves inward as stellar mass decreases \citep[$a_{\rm snow}
\propto M_\star^{1-2}$, e.g.
][]{2008ApJ...673..487I,2008ApJ...673..502K}, so scattering remains
difficult. For $a_{\rm snow} \propto M_\star$, $v_K(a_{\rm snow})$ is
constant for different stellar masses. However, for a fixed time
period, a greater number of conjunctions for low-mass stars allows
dynamical evolution to greater eccentricities.

\subsection{Scattering Simulations}\label{sec:sim-scattering}

To measure the likelihood of planet-planet scattering, we performed
simulations over a range of stellar masses with the MERCURY integrator
\citep{1999MNRAS.304..793C}. We initialised integrations with two
10\,$M_\oplus$ planets spaced near the Hill stability criterion to
ensure close encounters \citep{1993Icar..106..247G}. This planet mass
is an approximate maximum mass before cores accrete gas to become gas
giants, and thus offers the best chance for scattering over
coalescence. To represent a linearly stellar mass dependent snow line,
the inner planet was placed at $a_{\rm in} = 3\,M_\star$\,AU. The
outer planet begins at a random $a$ in the range 0.9--$1 \, a_{\rm in}
\left( 1 + \Delta_{\rm crit}\right)$, where $\Delta_{\rm crit} = 3
\left( M_{\rm pl}/M_\star \right)^{1/3}$
\citep{1993Icar..106..247G,2007astro.ph..3163F}. Both planets begin in
circular orbits with random inclinations less than 3$^\circ$; the
remaining orbital elements are chosen randomly. Simulations were run
with a 5\,day timestep for 1\,Gyr around stars of 0.25, 0.5, 1, and
2\,$M_\odot$, or halted earlier in the case of collisions (we assume
perfect mergers) or ejections. A total of 520 simulations were run,
130 for each stellar mass.

\subsection{Scattering Results}\label{sec:results-scattering}

The simulations result in three different outcomes: collisions,
ejections, or survival of both planets for 1\,Gyr. No planets achieved
periastra low enough to fall onto the central star. Most ($>85\%$)
simulations resulted in collisions (Table \ref{tab:scattering}). Some
systems survived for the full simulation. The only ejections were for
0.25\,$M_\odot$.

\begin{deluxetable}{cccc|c}
  \tablecolumns{4} \tablecaption{Scattering simulation outcomes, and
    fraction with low periastra\label{tab:scattering}.}
  \tablewidth{0pt} \tablehead{\colhead{$M_\star(M_\odot)$} &
    \colhead{collisions} & \colhead{ejections} & \colhead{survival} &
    \colhead{$q < 0.1$\,AU} } \startdata
  0.25 & 95\% & 2.5\% & 2.5\% & 4\% \\
  0.5  & 95\% & 0\%   & 5\%   & 0\% \\
  1    & 94\% & 0\%   & 6\%   & 0\% \\
  2    & 84\% & 0\%   & 16\%  & 0\% \\
\enddata
\end{deluxetable}

\begin{figure}
\begin{center}
  \plotone{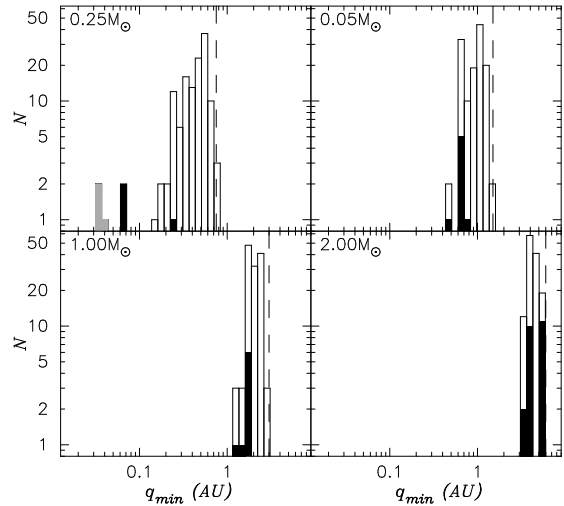}
  \caption{Smallest periastra of scattering simulations for all
    non-stable orbits. Black filled bars: simulations in which both
    planets survived until 1\,Gyr. Grey bars: ejections (orbits going
    beyond 1000\,AU). Unfilled bars: collisions/mergers. The dashed
    line shows the starting $a$ (and $q$) of the inner planets
    (circular) orbit. Planets are scattered to lower periastra for
    low-mass stars because of more conjunctions, and starting closer
    to the central star.}\label{fig:rfhist}
\end{center}
\end{figure}

With so few systems remaining after 1\,Gyr, we use the smallest
periastron distance reached in each simulation to characterise the
success of planet-planet scattering, shown in Figure
\ref{fig:rfhist}. As expected, the closer snow line distance for the
0.25\,$M_\odot$ allows smaller periastra after scattering.

For simulations of 0.25\,$M_\odot$ stars, 5/130 (4\%) planets reach
periastra less than 0.1\,AU. A shorter orbital period allows many more
conjunctions. Thus, systems evolve further than for more massive
stars. For the three ejections, the lowest periastra were reached just
before a series of close encounters, which resulted in the
ejection. For the three surviving systems, the lowest periastra were
reached near the end of the integrations. These orbits have
eccentricities $\approx$0.5, and semi-major axes $\approx$0.5\,AU,
corresponding to an orbital period of around
260\,days. Circularisation times for these planets are $\sim$10\,Gyr
\citep{1966Icar....5..375G,2007arXiv0711.2015R}.

Therefore, in the case of 10\,$M_\oplus$ planet-planet scattering,
only the lowest mass stars have planets with periastra close enough
for transiting orbits. However, long circularisation timescales mean
these planets will likely remain on highly eccentric orbits, with
periods long enough to make radial velocity and transit detections
difficult.

\section{Migration}\label{sec:migration}

We now turn to a migration scenario. In this picture, the largest
protoplanets form near the snow line by oligarchic growth. Once they
reach masses of $\sim$1--10\,$M_\oplus$, these icy protoplanets
migrate towards the central star. In a cold finger disk, migrating icy
objects dominate smaller interior rocky protoplanets. Long chaotic
growth timescales mean that as the icy object migrates through the
terrestrial region, interior rocky objects are not accreted. They are
instead scattered outward or shepherded inward. Shepherding---where
interior objects are captured into mean-motion resonances---results in
rocky protoplanets being pushed inward ahead of the migrating icy
planet. These smaller objects merge to form large rocky planets, which
are eventually accreted by the larger icy migrator or survive on an
interior orbit. We assume that all objects halt their migration when
they reach the inner edge of the gas disk, at $\sim$10 stellar radii.

Our goal is to calculate the growth and migration of individual
protoplanets in this scenario. If several protoplanets migrate,
whether they do so as a set in resonant orbits
\citep{2007ApJ...654.1110T}, or successively
\citep[e.g.][]{2006Icar..185..492D,2008ApJ...673..487I}, the resulting
trends are similar in our picture. The main difference between
outcomes is the number of icy planets on short-period orbits. The
trends our models predict depend largely on disk and planet
properties, not multiplicity.

When an icy protoplanet migrates, it only interacts dynamically with
interior objects. Because collision cross sections are essentially
geometric, the timescale for growth is much longer than the migration
timescale. For example, the migration timescale for an Earth-mass
planet at 1\,AU (several 10$^5$\,yr, or several 10$^6$\,yr if
migration is less efficient) is much smaller than the chaotic growth
timescale \citep[$\tau_{\rm chaotic} \sim \rho R_{\rm pl} / \sigma_s
\Omega \sim 10^8$\,yr, where $R_{\rm pl}$ is the planet
radius,][]{2004ApJ...614..497G}. Thus, the migrating protoplanet does
not accrete terrestrial protoplanets; outward scattering or inward
shepherding are the most likely outcomes.

The evolution of interior protoplanets depends on their random
velocities. Chaotically growing objects with high eccentricities are
scattered outward by the migrating protoplanet. These may interact
with another migrating protoplanet or resume chaotic growth. If
interior objects have finished chaotic growth and are damped by the
gas disk onto more circular orbits, shepherding by capture onto
resonant orbits is possible. Shepherded objects merge and form rocky
planets as their orbits are pushed together by the migrating icy
protoplanet. Shepherding by giant planets undergoing type II migration
has been proposed as a way to form super Earth-mass planets
\citep{2005ApJ...631L..85Z}. However, studies have yet to consider
shepherding by super-Earths undergoing type I migration.

While some planets are stranded at intermediate radii as the gas disk
dissipates, most planets that begin to migrate reach the inner disk
edge, and might fall onto the star. Because the torque on the
migrating planet changes when the disk gas surface density profile
varies rapidly, as happens at the inner disk edge, this fate may be
avoided \citep{2002ApJ...565.1257T}. Here, corotation torques affect
migration, and allow for planets to cease migration before reaching
the stellar surface \citep{2006ApJ...642..478M}. In our migration
simulations, we therefore assume migration stops inside the inner disk
edge \citep{2007ApJ...654.1110T}.

In the rest of this section, we consider three models that explore
different aspects of the migration scenario, and observable trends
that probe stellar and disk properties. We consider the simplest
scenario---when growth is so fast that planets reach isolation before
migration begins---with an analytic model in \S \ref{sec:analytic}. As
the planetesimal size increases, growth slows; the timescale becomes
comparable to that for migration. The assumption made in the analytic
model no longer applies, and we use a semi-analytic model to study
concurrent growth and migration in \S \ref{sec:model}. Finally, we use
$n$-body simulations in \S \ref{sec:shepherding} to show the
shepherding effects migrating super-Earths have on terrestrial
material.

\subsection{An Analytic Approach}\label{sec:analytic}

If we assume that protoplanets reach isolation before migration
starts, then we can create a simple analytical model for our migration
scenario. At isolation, protoplanets have a known migration timescale,
which is shorter than the disk lifetime if they are to reach the
central star. To remain in the super-Earth mass regime, the mass of a
protoplanet is smaller than the critical core mass for gas
accretion. Because the isolation mass changes with surface
density---and thus with disk mass---only a certain range of disk
masses satisfy these conditions for fixed stellar mass. To consider a
range of different stars, we also consider how the snow line---where
these migrating planets form---changes with stellar mass. The range of
disk masses that satisfy the conditions changes with stellar mass,
resulting in potentially observable trends that test migration models.

To begin, we adopt a relation for the surface density of solid
material in the disk. In the standard MMSN model,
\begin{equation}
  \sigma_{\rm s} = \sigma_0 \, f_{\rm ice} \, a_{\rm AU}^{-\delta} \, ,
\end{equation}
where $\sigma_0 = 8$\,g cm$^{-2}$, $\delta = 1$--1.5, and $a_{\rm AU}$
is $a$ in units of AU. The factor $f_{\rm ice} \sim 2$--3 is the
enhancement from ice condensation beyond the snow line. This disk has
a mass $\sim$0.01\,$M_\odot$.

To generalise this relation, we add terms to account for differences
in disk mass and metallicity around stars with a range of masses.
Disks around young stars have a large dispersion in mass
\citep{2000prpl.conf..559N,2005ApJ...631.1134A,2007ApJ...671.1800A}. Setting
the disk mass $M_{\rm disk} \propto \eta M_\star^\beta$ allows us to
treat the observed trends with stellar mass---$M_{\rm disk} \propto
M_\star^\beta$, with $\beta \approx 1$---and a range ($\eta$) of disk
masses at fixed stellar mass. Adopting a factor $\mathcal{M} \propto
10^{\rm [Fe/H]}$ for the metallicity of the stars and the disk yields
\begin{equation}\label{eq:sigma}
  \sigma_{\rm s} = \sigma_0 \, \eta f_{\rm ice} \, \mathcal{M} \, M_\star^\beta
  \, a_{\rm AU}^{-\delta} \, .
\end{equation}
For simplicity, we combine $f_{\rm ice}$ and $\mathcal{M}$ into a
single factor $\Delta = f_{\rm ice} \mathcal{M}$, which quantifies the
enhancement of solid material relative to gas where these planets
form. For a cold finger disk, we use $f_{\rm ice} = 10$. Thus, for
typical ranges in $\mathcal{M}$ ($\sim$1/3--3) and $f_{\rm ice}$
(2--10), the plausible range of $\Delta$ is 0.6--30. We concentrate on
higher $\Delta$, because these are cold finger disks.

For the surface density of the gas disk, we set $\sigma_g = 100
\sigma_s / \Delta$. Thus, the gas mass depends on $\eta$ and $\delta$,
and is independent of metallicity and the enhancement in ices at the
snow line. We adopt $\delta = 3/2$.

How the snow line varies with stellar mass is uncertain. The existence
of gas giant planets suggests that the stages of planet formation up
to isolation occur while the gas disk is still present. During these
stages the snow line distance is set by viscous accretion of the gas
disk. If the accretion rate onto the star is $\dot{M} \propto
M_\star^{1-2}$, then $a_{\rm snow} \propto M_\star^{6/9 - 8/9}$
\citep{2008ApJ...673..502K}. Later, when the star has reached the
main-sequence and the gas disk has been dissipated, the main-sequence
luminosity is more important and $a_{\rm snow} \propto M_\star^2$
\citep{2005ApJ...626.1045I}. Because we model oligarchic growth, and
$\dot{M} \propto M_\star^2$ \citep{2005ApJ...625..906M}, we adopt the
snow line distance $a_{\rm snow} = 2.7 M_\star$\,AU. Variation of the
snow line with time and stellar mass is a key component of planet
formation models that consider a range of spectral types
\citep{2008ApJ...673..502K}.

Substituting our adopted surface density into the isolation mass
yields
\begin{equation}\label{eq:miso3}
  M_{\rm iso} \propto \frac{ \left( \sigma_{\rm s} \, a^2 \right)^{3/2} }
  { \left( M_\star \right)^{1/2} } =
  \frac{ \left( \eta \, f_{\rm ice} \, \mathcal{M} \, M_\star \,
      a^{1/2} \right)^{3/2} }
  { M_\star^{1/2} } \, .
\end{equation}
The isolation mass increases with any parameter that increases the
surface density. The increasing disk mass with stellar mass ($M_\star$
in numerator) is stronger than the decreasing Hill radius ($M_\star$
in denominator). Thus, at fixed $a$ the isolation mass increases with
stellar mass. For our scenario, we are interested in planets that form
at the snow line, so the changing snow line distance ($a = a_{\rm
  snow} \propto M_\star$) makes the stellar mass dependence stronger.

Substituting $a = 2.7 M_\star$\,AU, equation (\ref{eq:miso3}) yields
the isolation mass \emph{at the snow line} for a range of stellar and
disk masses, and metallicities and snow line enhancements
\begin{equation}\label{eq:miso}
  M_{\rm iso} = 0.12 \, \left( \Delta \, \eta \right)^{3/2}
  M_\star^{7/4} M_\oplus \, .
\end{equation}

Applying the same approach to type I migration yields
\begin{equation}\label{eq:taumig1}
  \tau_{\rm mig} \propto \frac{ M_\star^2 \, h^2 \, f_{\rm mig} }
  { a^2 \, M_{\rm pl} \, \Omega \, \sigma_{\rm gas} } =
  \frac{ M_\star^{1/2} \, h^2 \, f_{\rm mig} \, a }{ M_{\rm pl} \, \eta } \, ,
\end{equation}
where the offset $f_{\rm mig}$ allows us to consider reduced migration
rates. At fixed $a$, migration takes longer as stellar mass increases,
and speeds up as planet mass increases. If planet masses vary less
strongly with radial distance than $M_{\rm pl} \propto a$, then the
migration timescale increases outward, and planets cannot catch up to
interior ones. Even with isolated objects ($M_{\rm iso} \propto
a^{3/2}$), planets may not catch up to interior ones due to the strong
slowing of growth with semi-major axis. At the snow line distance,
migration slows even more strongly with increasing stellar mass due to
lower gas density and slower orbital periods at larger radii.

Again substituting $a = 2.7 M_\star$\,AU, the timescale to migrate
from the snow line to the star is
\begin{equation}\label{eq:tmig}
  \tau_{\rm mig} = 9.1 \times 10^5 \,
  \frac{ f_{\rm mig} \, M_\star^{3/2} }{ M_{\rm pl} \, \eta } \, {\rm yr}
  \, ,
\end{equation}
where we have set $h = 0.05$ \citep[e.g.][]{2007prpl.conf..655P}. At
fixed $M_\star$, massive planets in massive disks migrate to the inner
disk edge fastest. The migration timescale increases with $M_\star$
because the snow line is further away.

If the migration time is shorter than the disk lifetime
(i.e. $\tau_{\rm mig} \lesssim \tau_{\rm disk} \sim 1$\,Myr), then
protoplanets reach short-period orbits. This inequality leads to
\begin{equation}\label{eq:miso2}
  M_{\rm pl} > \frac{ 0.91 \, f_{\rm mig}
    \, M_\star^{3/2} }{ \eta } \, M_\oplus \, .
\end{equation}
This result yields the minimum mass for a planet to migrate to a close
orbit. Substituting the isolation mass (eq. \ref{eq:miso}) for $M_{\rm
  pl}$ and solving for $\eta$ gives a lower relative disk mass limit
of
\begin{equation}\label{eq:llim}
  \eta > \eta_{\rm low}
  = \frac{ 2.2 \, f_{\rm mig}^{2/5} }{ M_\star^{1/10} \, \Delta^{3/5}
  } \, .
\end{equation}
Disks more massive than this $\eta$ form protoplanets massive enough
to migrate to short-period orbits before the gas disk
dissipates. Planets in slightly less massive disks still migrate, but
are stranded at intermediate radii as the disk disperses.

The critical $\sim$10\,$M_\oplus$ core mass for gas accretion provides
an upper limit for the protoplanet mass. Solving $M_{\rm iso} <
10\,M_\oplus$ for $\eta$ yields
\begin{equation}\label{eq:ulim}
  \eta < \eta_{\rm hi} = \left\{ \begin{array}{ll}
        \frac{ 18.6 }{ M_\star^{7/6} \, \Delta } &
        \frac{ 18.6 }{ M_\star^{7/6} \, \Delta } < 30 \\
        30 &  \frac{ 18.6 }{ M_\star^{7/6} \, \Delta } \geq 30
      \end{array}
    \right. \, ,
\end{equation}
where the additional constraint of a reasonable disk mass sets $\eta
\lesssim 30$ ($M_{\rm disk} \lesssim 0.3\,M_\star$) as an upper limit
\citep[e.g.][]{2005ApJ...626.1045I}. Because we assume growth is fast,
planetesimal accretion drops significantly at later stages. The core
mass for gas accretion is then somewhat smaller
\citep{2000ApJ...537.1013I,2006ApJ...648..666R}.

The two limits on disk mass yield a simple relation between the
stellar mass, migration offset, and enhancement factor. Equating
$\eta_{\rm low}$ and $\eta_{\rm hi}$,
\begin{equation}\label{eq:mstarmax}
  M_{\star,{\rm max}} = \frac{ 7.3 }
  { \left( f_{\rm mig} \, \Delta \right)^{3/8} }
\end{equation}
in units of Solar masses.\footnote{Equating (\ref{eq:llim}) and
  (\ref{eq:ulim}) has two solutions for $M_\star$ because of the upper
  limit of 30. The other solution is at $M_\star$ far too small to be
  interesting.} This equation has a simple physical
interpretation. For massive stars ($M_\star > M_{\star,{\rm max}}$),
the only protoplanets massive enough to migrate to the central star
before the gas disk disperses are above the critical core mass for gas
accretion. These planets therefore become gas giants, rather than
forming hot super-Earths. For lower stellar masses, the closer snow
line distance allows planets smaller than the critical core mass to
migrate to the host star. Thus, $M_{\star,{\rm max}}$ is the maximum
stellar mass for hot super-Earths produced by type I migration.

Making an estimate of $M_{\star,{\rm max}}$ requires an assumed
$f_{\rm mig}$ and $\Delta$. For Solar metallicity $\mathcal{M} = 1$,
and a cold finger enhancement $f_{\rm ice} = 10$--20, $\Delta =
10$--20. For a migration offset $f_{\rm mig} = 10$, $M_{\star,{\rm
    max}} \sim 1\,M_\odot$. Transit and radial velocity surveys
routinely probe these stellar masses. Independent of the disk mass
distribution, this result is therefore a simple testable prediction of
hot super-Earth formation by type I migration.

\begin{figure}
\begin{center}
  \plotone{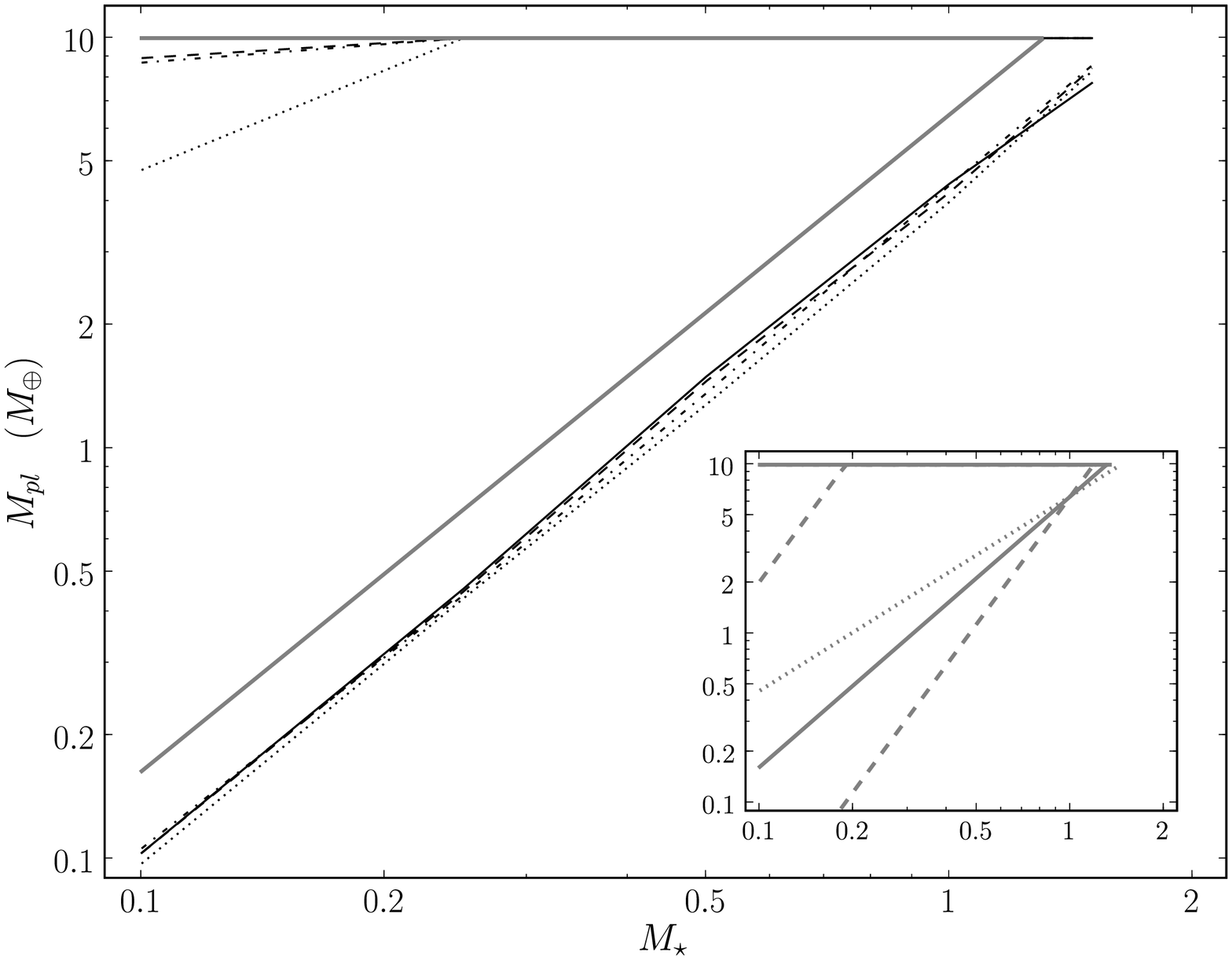}
  \caption{The range of planet masses that reach close-in orbits as a
    function of stellar mass. The thicker grey line line shows results
    from the analytic model using equations (\ref{eq:miso}),
    (\ref{eq:llim}), and (\ref{eq:ulim}). Other lines show upper and
    lower mass limits for a range of planetesimal sizes from the
    semi-analytic model (see \S \ref{sec:model}) for $\Delta = 10$: $r
    = 10$\,m (plain), 100\,m (dashed), 1\,km (dot-dashed), and 10\,km
    (dotted). The range of planet masses reaching short-period orbits
    decreases with increasing stellar mass because the snow line
    distance is greater. The inset panel (same axes) shows how
    different snow line relations affect the model (using $a_{\rm
      snow} = 2.7 M_\star^\alpha$\,AU). Lines are for $\alpha = 1/2$
    (dotted), 1 (plain), and 2 (dashed) A more strongly varying snow
    line distance ($\alpha = 2$) yields smaller $M_{\rm iso}$ (due to
    smaller $R_{\rm H}$) at much closer snow line distances as stellar
    mass decreases.}\label{fig:f10sum}
\end{center}
\end{figure}

\begin{figure*}
\begin{center}
  \plottwo{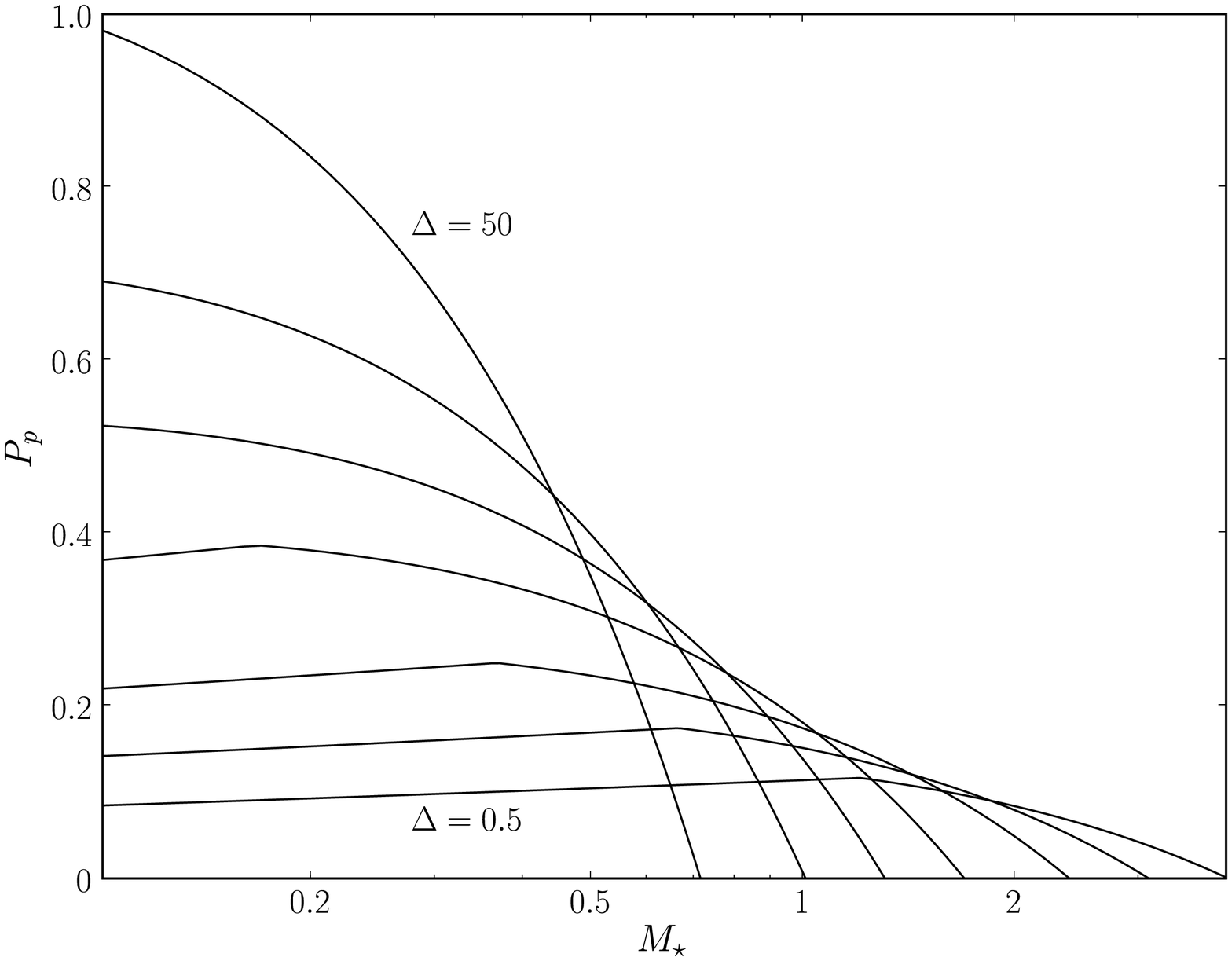}{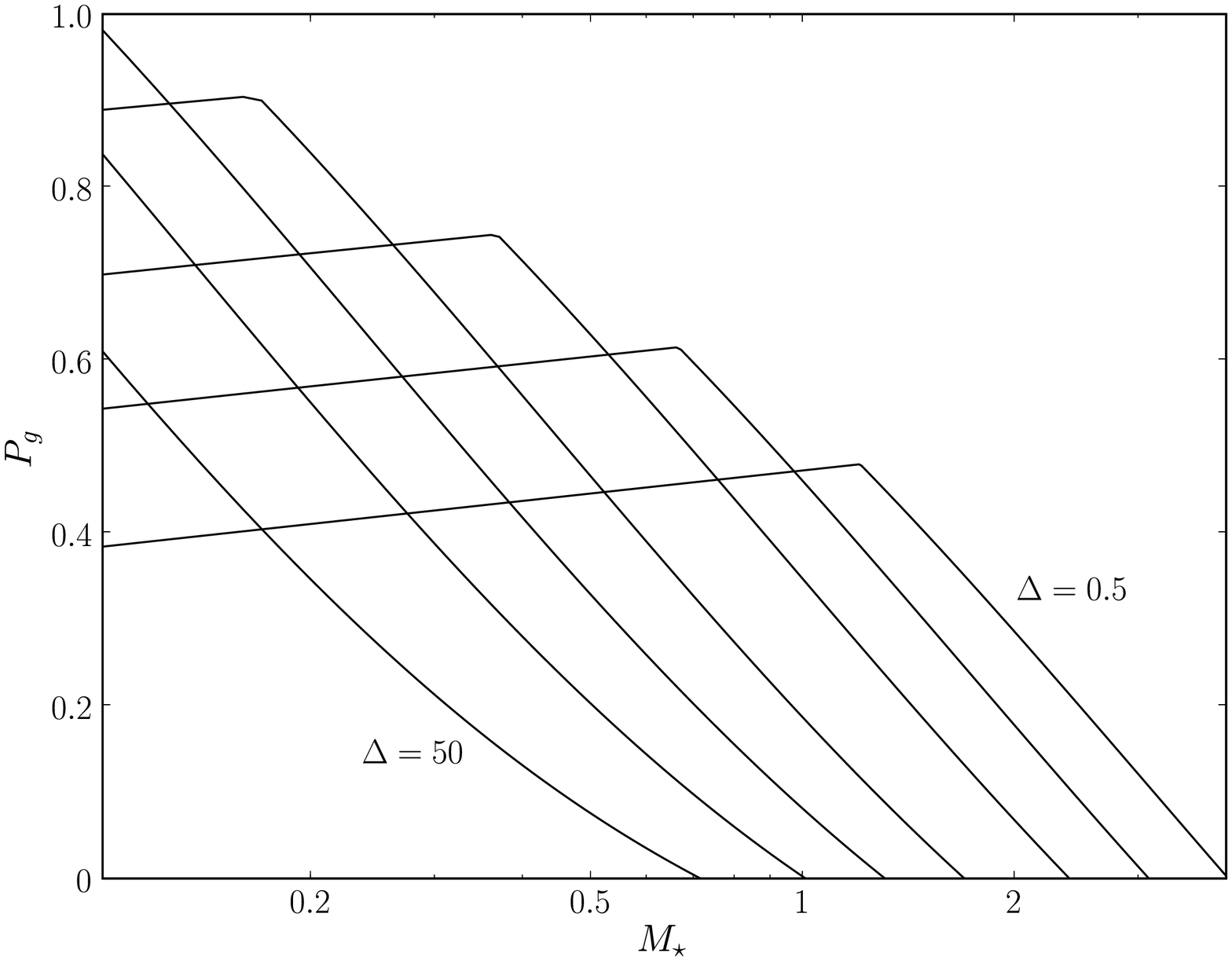}
  \caption{Probability distributions for power law (left), and
    Gaussian with $\mu = 1$ (right) disk mass distributions of
    $\lesssim$10\,$M_\oplus$ planets as a function of stellar
    mass. For $f_{\rm mig} = 10$, lines are for $\Delta = 0.5$, 1, 2,
    5, 10, 20, and 50 from right to left where curves cross the
    x-axis. Thus $M_{\star,{\rm max}}$ decreases with increasing
    $\Delta$, and is independent of the disk mass distribution). Each
    plot is arbitrarily normalised to 1 for the most likely $\Delta$
    at $M_\star = 0.1\,M_\odot$.}\label{fig:prob}
\end{center}
\end{figure*}

Figure \ref{fig:f10sum} shows the range of planet masses that reach
short-period orbits for a range of stellar masses. For the analytic
model (thick grey lines) the upper limit is constant at
10\,$M_\oplus$. The lower limit decreases as stellar mass and snow
line distance decrease. The expected range of planet masses decreases
with increasing stellar mass, while the average mass increases to
10\,$M_\oplus$, where the lines meet at $M_{\star,{\rm max}} =
1.3$\,$M_\odot$.

In addition to this maximum stellar mass, we can derive the
probability of forming hot super-Earths around stars with $M_\star <
M_{\star,{\rm max}}$. This estimate requires an adopted distribution
of $\eta$ (i.e. disk masses). If relative disk masses ($M_{\rm
  disk}/M_\star$) are distributed as a power law with index
$\sim$$-1.75$ \citep{2005ApJ...631.1134A}, the (relative) probability
of forming a close-in planet as a function of stellar mass for a
given $\Delta$ is
\begin{equation}\label{eq:pprob}
  P_{\rm p}(M_\star,\Delta) \propto
  \int_{\eta_{\rm lo}}^{\eta_{\rm hi}} \eta^{-1.75} \, d\eta \, .
\end{equation}
Alternatively, disk masses may be distributed around some ``typical''
relative disk mass \citep[e.g.][]{2005ApJ...626.1045I}
\begin{equation}\label{eq:gprob}
  P_{\rm g}(M_\star,\Delta) \propto \int_{\eta_{\rm lo}}^{\eta_{\rm hi}}
  \exp \left( - \frac{ \left( \log(\eta) - \mu \right)^2 }{ 2 \, s^2 } \right)
  d\eta
\end{equation}
where we choose the standard deviation $s = 1$. This distribution is
plausible because opacities may underestimate disk masses by as much
as an order of magnitude, due to mass locked up in boulder size
objects \citep{2007ApJ...671.1800A}. Therefore mm observations see
disks not only with a range of masses, but in a range of evolutionary
states. Unlike the case for giant planets, there is no observational
anchor point, so we present these results as relative probabilities.

For a range of $\Delta$, the left panel of Figure \ref{fig:prob} shows
the probability distribution for the power law disk mass distribution
with $f_{\rm mig} = 10$. Results are similar for $f_{\rm mig} = 1$,
with the main difference that $M_{\star,{\rm max}}$ is higher
(eq. \ref{eq:mstarmax}). Higher $\Delta$ are most relevant here
because low values describe MMSN disks, which result in many
similar-mass migrating planets originating from a wide range of
radii. The point where lines break and decrease toward lower stellar
masses is caused by the maximum disk mass condition $\eta < 30$. In
these cases the maximum short-period planet mass is not set by gas
accretion, and is $<$10$M_\oplus$. This limit applies when $\Delta
\lesssim 5$ for the lower of the stellar masses we consider, so does
not apply to cold finger disks with $f_{\rm ice} \gtrsim 10$ unless
they have metallicity $\mathcal{M} \lesssim 0.5$.

At the lowest stellar masses, there is a clear increase in planet
frequency with $\Delta$. With a power-law distribution of disk masses,
the most common disks are the least massive; these require large
$\Delta$ to allow them to form planets massive enough to migrate (and
satisfy condition \ref{eq:llim}). Near $M_{\star,{\rm max}}$, there is
an optimum $\Delta$, which is a balance between the likelihood of
different disk masses and the $\Delta$ needed to form close-in planets
from those disks. At $M_{\star,{\rm max}}$, the only planet that
reaches a short-period orbit has $M_{\rm pl} =
10\,M_\oplus$. Therefore the range of short-period planet masses
decreases up to $M_{\star,{\rm max}}$. The average planet mass
increases with stellar mass.

The right panel of Figure \ref{fig:prob} shows the probability
distribution for the Gaussian distribution with $\mu = 1$.  The most
common disk mass is thus $\sim$0.1\,$M_\star$. As $\Delta$ increases,
the probability of forming a short-period planet decreases once the
disk mass distribution is not truncated by the condition $\eta <
30$. In contrast to the power law distribution, the low-mass disks
requiring large $\Delta$ are uncommon. Thus, as $\Delta$ increases,
isolation masses are pushed over the gas accretion mass, and the
likelihood of forming close-in $\lesssim$10\,$M_\oplus$ planets
decreases. While the curves are different from the left panel, the
point $M_{\star,{\rm max}}$ is the same for a given $\Delta$.  With
$\mu = 0$ (i.e. distributed about $M_{\rm disk} = 0.01\,M_\star$) the
probability distribution is qualitatively similar to the power law
disk distribution.

In summary, the simple analytical model yields testable predictions
for an ensemble of super-Earths that migrate into short-period orbits
from the snow line. For reasonable input parameters, we predict a
maximum stellar mass $\sim$1\,$M_\odot$ for stars with close-in
super-Earths. If circumstellar disks tend to have similar snow line
enhancements, this maximum mass decreases with the metallicity of the
host star. For a range of stellar masses, the frequency of hot
super-Earths depends on the initial distribution of disk masses. For a
power-law (Gaussian) distribution of disk masses, the model predicts
more (fewer) hot super-Earths around more metal-rich stars.

To give the these trends some context, the first transiting low-mass
planet orbits a star with sub-Solar mass and metallicity
\citep[GJ436b,][]{2007A&A...472L..13G}. The current sample of low
minimum-mass planets also indicates a flatter metallicity distribution
than exists for giant extra-Solar planets
\citep{2007prpl.conf..685U}. While both disk mass distributions
suggest that low stellar mass host is likely, the power law
distribution argues against a low metallicity host. The Gaussian disk
mass distribution, centered on a relatively high disk mass is
consistent with an increasing giant planet frequency with metallicity,
and a flatter or decreasing frequency for lower mass planets.

Disks with $\eta > \eta_{\rm hi}$ form gas giants. Their relative
probabilities can thus be calculated by integrating equations
(\ref{eq:pprob}) and (\ref{eq:gprob}) from $\eta_{\rm hi}$ to
30. However, because $\eta_{\rm low}$ only weakly depends on $M_\star$,
giant planet frequency is roughly some constant minus the hot
super-Earth frequency (i.e. generally increases with $M_\star$). This
trend is essentially the result arrived at by previous theoretical
studies \citep[e.g.][]{2005ApJ...626.1045I,2008ApJ...673..502K}, and
is at least qualitatively consistent with the observed trend
\citep{2007ApJ...670..833J}.

In constructing the above model we simplified some parameters, and
assumed values for others. We now briefly consider model sensitivity
to these, and whether observations may constrain them. The most
uncertain simplification is how the snow line distance varies with
stellar mass.  Within our framework, relaxing the distance to $a_{\rm
  snow} = 2.7 M_\star^\alpha$\,AU results in changes to Equations
(\ref{eq:ulim}), (\ref{eq:llim}), and (\ref{eq:mstarmax}) for $\alpha
= 1/2$--2 (Fig. \ref{fig:f10sum} inset). A more strongly varying snow
line distance ($\alpha = 2$) yields much closer $a_{\rm snow}$ and
smaller $M_{\rm iso}$ (due to smaller $R_{\rm H}$) for low mass
stars. A more complex snow line model could include how $a_{\rm snow}$
varies with $M_{\rm disk}$ at fixed stellar mass, or some time
dependence \citep[e.g.][]{2006Icar..181..178C,2008ApJ...673..502K}.

Another uncertain parameter is $\delta$, the disk surface density
power-law index. While we used $\delta = 3/2$, many models also
consider $\delta = 1$. With $\delta = 1$, the main results of Figure
\ref{fig:f10sum} are unchanged, with stronger migration accounting for
lower mass planets as the snow line distance decreases. It is unlikely
observations of short-period super-Earths can constrain $\alpha$ or
$\delta$ based on Figure \ref{fig:f10sum}, because they affect lower
limits to planet masses, which will be hard to detect.

The efficiency of type I migration is also unclear. Our choice of
$f_{\rm mig} = 10$ is based on numerical simulations, but may also be
probed by future discoveries. The maximum stellar mass $M_{\star,{\rm
    max}}$ is not very sensitive to the snow line distance or disk
profile, so for fixed snow line and metallicity enhancements
($\Delta$), observations probe values for $f_{\rm mig}$.

Our final major assumption is that planets form rapidly, and reach
isolation before migrating. If planetesimals are small and growth is
shear dominated, this assumption is generally true. With larger
planetesimals however, growth is slower and planets may leave their
formation regions while still growing. Planetary growth and migration
are then coupled, and must be calculated simultaneously. Recently,
\citet{2006Icar..180..496C,2006ApJ...652L.133C} showed how a
semi-analytic model of oligarchic growth can take different
planetesimal sizes into account, and estimate their effect on growth
rates \citep[see also][]{2003Icar..161..431T,2008Icar..194..800B}. We
now turn to a similar, yet simplified model to estimate the effects of
planetesimal size on growth and migration.

\subsection{Semi-Analytic Model}\label{sec:model}

If planets grow fast enough, the isolation mass sets the range of disk
masses that form migrating planets. If planetesimals are large enough,
growth is not shear dominated and is slower. Migration then begins
before planets reach isolation. To follow this evolution, a model
treating concurrent accretion and migration is necessary. Our model
tracks damping of planetesimal random velocities by gas drag and
stirring by a growing protoplanet. The random velocities set how
growth proceeds relative to migration, allowing comparison with the
analytic model.

In the model, a single protoplanet of mass $M_{\rm pl}$ grows on a
circular orbit from a planetesimal disk of small bodies of radius $r$.
We adopt the accretion rate of \citet{2001Icar..149..235I} with the
atmosphere enhanced accretion radius of
\citet{2003A&A...410..711I}. To account for accretion of other nearby
protoplanets, the growth rate is increased by 50\%
\citep{2006ApJ...652L.133C}. Planetesimal random velocities are
stirred by the growing protoplanet \citep{2002Icar..155..436O} and
damped by gas drag \citep{2001Icar..149..235I}. The protoplanet
accretes and stirs material within an annulus of half-width 4\,$R_{\rm
  H}$, and undergoes type I migration at the rate derived by
\citet{2002ApJ...565.1257T}, modified by the offset $f_{\rm mig}$. We
use a ten times less efficient migration rate, motivated by numerical
\citep{2002A&A...385..647D,2003ApJ...586..540D,2006ApJ...652..730M},
and Monte-Carlo simulations \citep{2008ApJ...673..487I}. Objects have
mass density $\rho = 1.5$\,g cm$^{-3}$ outside the snow
line. Simulations are started with planetesimals in an equilibrium
between protoplanet stirring and gas drag. Because we consider growth
only near the snow line (see below), planetesimals do not undergo
radial motions due to gas drag. Planetesimals lost to gas drag can be
returned to the growth region by the cold finger mechanism
\citep{2004ApJ...614..490C}. The system is evolved using 4th order
Runge-Kutta integration with an adaptive step-size
\citep{1992nrca.book.....P}.

As before, we model protoplanets that form just outside the snow line.
These are the largest objects that migrate to the central star in a
cold finger disk and are largely unaffected by interior
objects. However, a migrating protoplanet shepherds material inward as
it migrates, and will accrete some terrestrial material. This
accretion cannot be treated by the semi-analytic model, so
protoplanets cease accretion once they pass inside the snow line in
the semi-analytic model. We model shepherding with $n$-body
simulations in \S \ref{sec:shepherding}. We vary $\eta$ to form
1--10\,$M_\oplus$ planets and use $\Delta = 10$.

Protoplanets begin with masses $1 \times 10^{-4}\,M_\oplus$, at
4\,$R_{\rm H}$ outside the snow line. This starting condition allows
them to reach isolation if growth is faster than migration. The disk
is split into 1000 equally spaced radial bins. However, because
accretion inside the snow line is turned off, objects grow from
material in $\sim$100 bins outside the snow line.

The snow line distance and gas disk are as in \S \ref{sec:analytic}
(eq. \ref{eq:sigma} and following text), but the surface density of
the gas disk decays exponentially with an e-folding time of 1\,Myr. We
place the inner edge of our disk at 0.2\,$M_\star$\,AU, though planets
that reach a few tenths of an AU are migrating so rapidly that the
exact value matters little.

To test our code, we compare growth at 5\,AU with Figure 1 from
\citet{2006ApJ...652L.133C}. His figure compares isolation times for
different $r$ with the type I migration timescale. The smallest size
planetesimals allow protoplanets to reach isolation before migration
starts. Growth was simulated at 5\,AU around a Solar-mass star, with a
solid surface density of 10\,g cm$^2$ (so $M_{\rm iso} \approx
10\,M_\oplus$), and a gas/solids ratio of 90. Migration was not
included, and the isolation time was simply compared to the analytic
estimate of Equation (\ref{eq:taumig}).

\begin{figure}[t]
\begin{center}
  \plotone{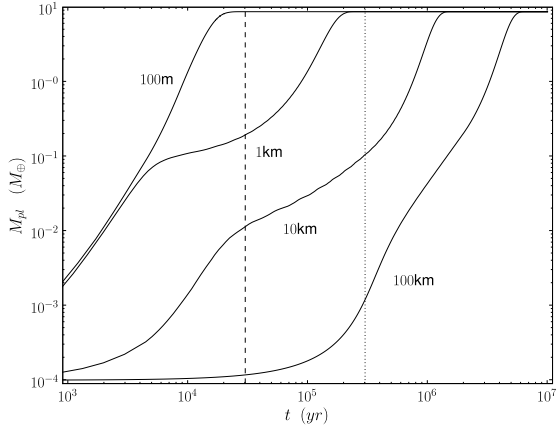}
  \caption{Growth to isolation of a protoplanet at 5\,AU around a
    Solar-mass star. Isolation times are in good agreement with Figure
    1 of \citet{2006ApJ...652L.133C}. Both models have $\sigma_0 =
    10$\,g cm$^{-2}$ and $\sigma_{\rm g} = 900$\,g cm $^{-2}$. Our
    model includes explicit calculation of planetesimal eccentricities
    and inclinations, which accounts for differences. Lines are for $r
    = 100$\,m, 1\,km, 10\,km, and 100\,km from left to right. The
    dashed (dotted) lines show the type I migration timescale for
    $f_{\rm mig} = 1$ (10).}\label{fig:test}
\end{center}
\end{figure}

Figure \ref{fig:test} shows growth in the absence of migration at
5\,AU around a Solar-mass star for a range of $r$ with similar initial
conditions. The time to reach isolation is fastest for the smallest
$r$ (100\,m), because growth is always shear dominated. For $r =
1$\,km, the growing protoplanet excites the small body random
velocities. Growth ceases to be shear dominated at several
$10^3$\,yr. For higher $r$, isolation takes even longer, due to the
decreasing effectiveness of gas drag on larger planetesimals.
Compared with Figure 1 from \citet{2006ApJ...652L.133C}, the time to
reach 10\,$M_\oplus$ is in good agreement. Our explicit calculation of
eccentricities and inclinations accounts for differences in how growth
proceeds \citep[c.f. Fig. 3 of ][]{2006Icar..180..496C}.

Models with sufficiently small planetesimals reach isolation before
migration. With $f_{\rm mig} = 1$, $r \lesssim 100$\,m, and for
$f_{\rm mig} = 10$, $r \lesssim 1$\,km. Thus, even with a reduced
migration rate, protoplanets may still migrate before isolation.

\subsubsection{Semi-Analytic Model Results}\label{sec:results}

\begin{figure*}
\begin{center}
  \plotone{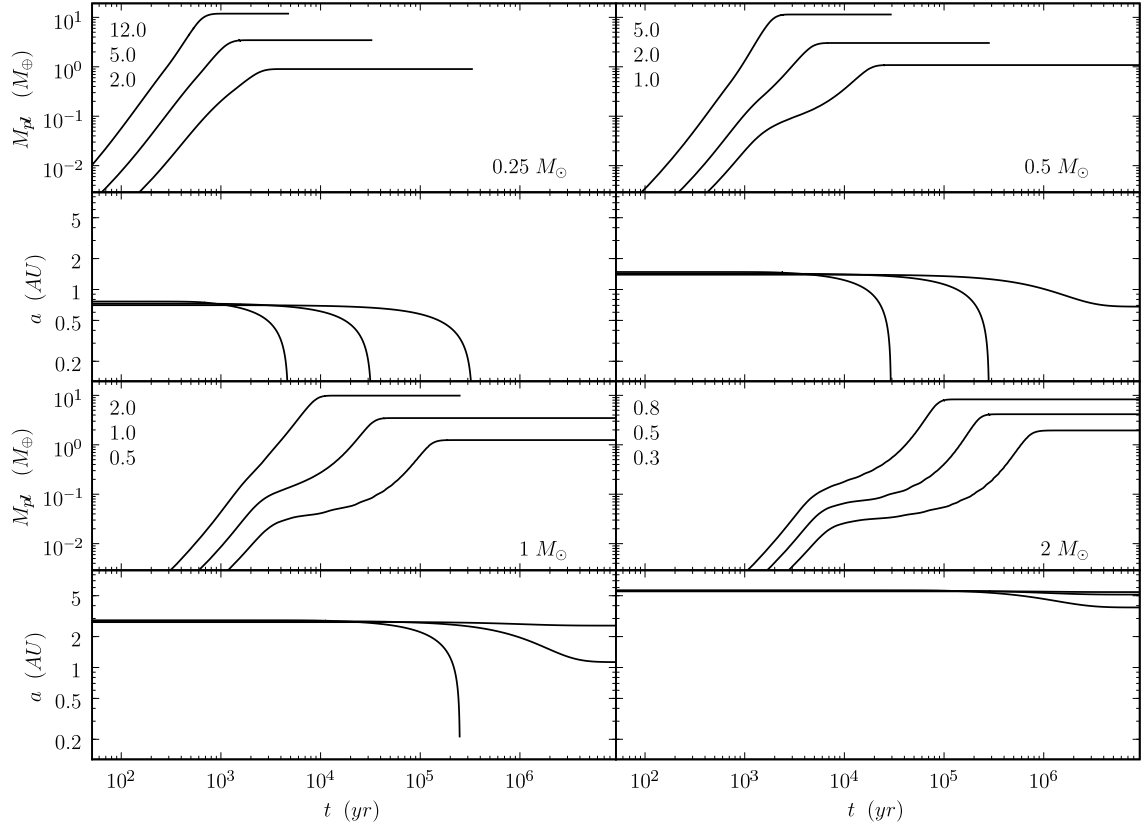}
  \caption{Results from the semi-analytic cold finger disk model with
    $\Delta = 10$ and $r = 100$\,m. Each vertical pair of panels shows
    mass and semi-major axis evolution for several relative disk
    masses. Higher disk masses ($\eta$ in legends) form more massive
    planets that migrate earlier. All 1--10\,$M_\oplus$ planets
    migrate to the inner disk edge for 0.25\,$M_\odot$, while none do
    for 2\,$M_\odot$.}\label{fig:f10r100m}
\end{center}
\end{figure*}

For the range of disk masses ($\eta$) that forms 1--10\,$M_\oplus$
planets around stars with masses 0.25--2\,$M_\odot$, Figure
\ref{fig:f10r100m} shows semi-major axis and mass evolution for $r =
100$\,m. The choice of 1\,$M_\oplus$ is somewhat arbitrary, but
represents a rough lower limit for detection. We first describe the
Solar case, and then look at differences as the stellar mass, and $r$
change.

For a Solar mass star, growth is not always fast enough for migration
to occur before the gas disk is dispersed. For $\eta = 2$, the objects
Hill radii increase faster than small bodies are stirred; thus growth
remains shear dominated ($\tilde{e},\tilde{i} \lesssim 1$). The
protoplanet successfully migrates to the inner edge of the disk. For
lower $\eta$, stirring overcomes damping at several $\times$10$^3$\,yr
and growth slows. Higher $\eta$ results in faster growth of larger
objects, which migrate early enough to avoid stalling at intermediate
radii. With $\eta = 1$, migration is somewhat significant, and the
$\sim$3 Earth mass planet stalls at $\sim$1\,AU due to dissipation of
the gaseous disk. For $\eta = 0.5$, the Earth-mass planet migrates
little, and remains beyond the snow line. Final planet masses, and the
degree of migration, are set by the initial surface density beyond the
snow line.

We turn now to trends across a range of stellar masses. Because of
smaller snow line distances, migration is easiest for planets in the
1--10\,$M_\oplus$ range around lower mass stars. Low-mass stars are
the most likely to form these planets, because the range of disk
masses that form them is much larger. For higher mass stars the more
distant snow line makes migration unlikely for all but the most
massive planets. Growth is driven out of the shear dominated regime
more easily due to lower gas density at greater distances. This result
confirms the maximum stellar mass $M_{\star,{\rm max}}$ described
above. As in the analytic model, $M_{\star,{\rm max}}$ lies between
1--2\,$M_\odot$ with $\Delta = 10$, because no planet with a mass
$\lesssim$10\,$M_\odot$ migrates significantly for 2\,$M_\odot$. As
stellar mass increases, the relative disk mass required to form
1--10\,$M_\oplus$ planets decreases (eq. \ref{eq:miso}).

\begin{figure*}
\begin{center}
  \plotone{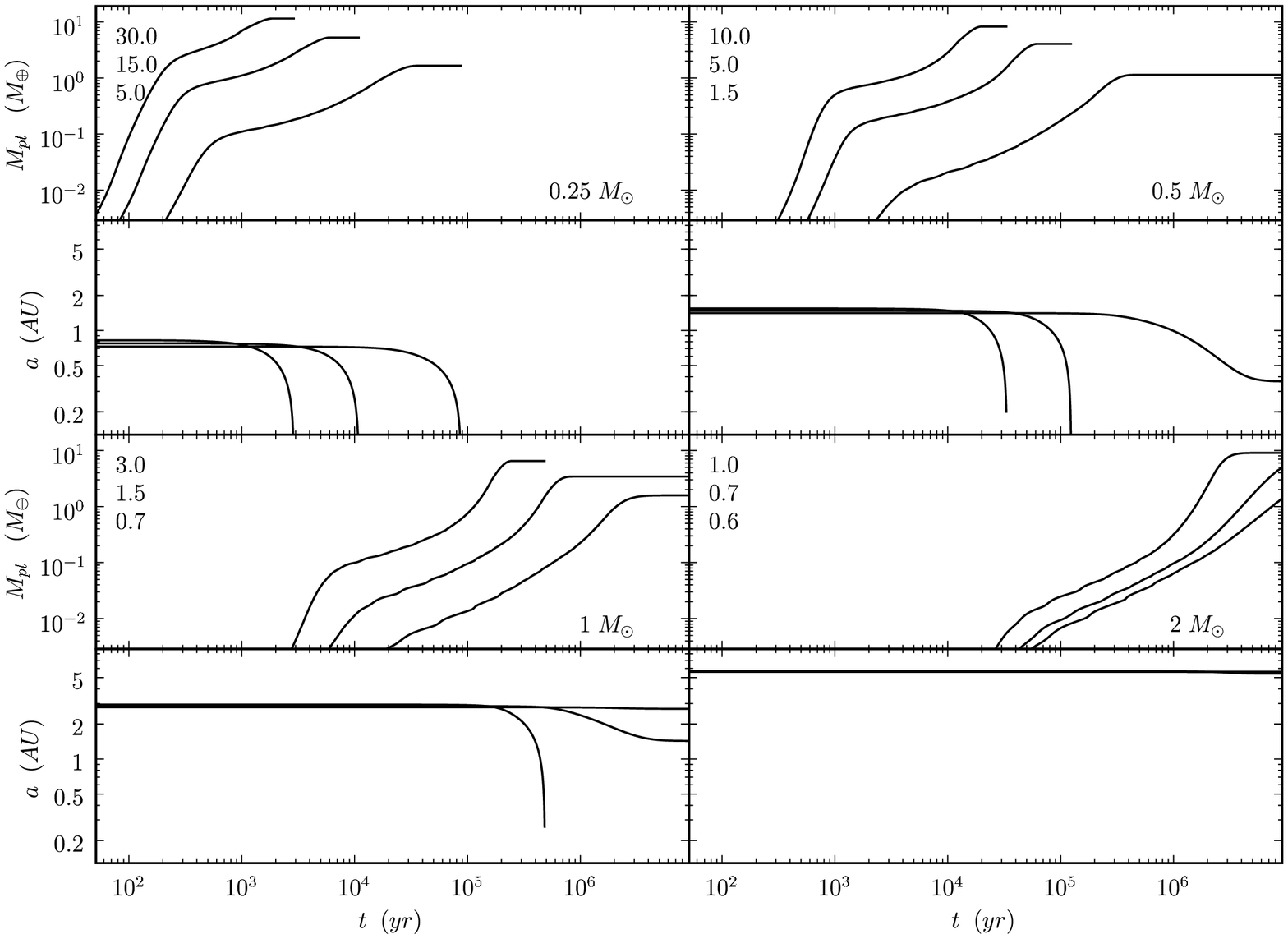}
  \caption{Same as Figure \ref{fig:f10r100m}, but with $r =
    10$\,km. Protoplanets still reach 1--10\,$M_\oplus$, but require
    much higher disk masses. Thus, results for protoplanets migrating
    to the inner disk edge are similar to the $r = 100$\,m
    case.}\label{fig:f10r10km}
\end{center}
\end{figure*}

Figure \ref{fig:f10r10km} shows how growth changes if planetesimals
are larger. Models again have $\Delta = 10$, but now the planetesimal
radius $r = 10$\,km. For larger planetesimals growth is easily stirred
out of the shear dominated regime by the large objects for all stellar
masses. The disk masses needed to reach the same range of planet
masses are higher, because planets migrate out of the accretion region
before they reach isolation. For 0.25\,$M_\odot$ stars, the maximum
$\eta = 30$ only just forms 10\,$M_\oplus$ planets. Again,
$M_{\star,{\rm max}}$ lies between 1--2\,$M_\odot$, indicating that it
is largely independent of planetesimal size. Though growth is slower,
the results for $r = 10$\,km are largely the same as 100\,m, because
the surface density can be increased to account for the slower growth.

Figure \ref{fig:f10sum} also includes results from the semi-analytic
model, showing the range of planet masses that reach short-period
orbits for a range of stellar masses. Models were run for $\Delta =
10$, with $r = 10$\,m--10\,km and $M_\star = 0.1$--2$M_\odot$. The
upper limit decreases at 0.1\,$M_\odot$ due to an upper limit on disk
masses. Results from the analytic model are in good agreement. The
difference in the lower limit arises because migration is faster at
smaller radii, allowing smaller planets to reach the inner disk edge
in the semi-analytic model.

In summary, using a more detailed migration model yields results
similar to the simple analytic treatment in \S \ref{sec:analytic}. The
inclusion of growth rates due to different planetesimal sizes adds
another dimension due to different relative timescales for migration
and accretion. The model offers more insight into how growth proceeds,
and how the growth rate sets the required disk mass for forming
short-period planets.

\subsection{Shepherding}\label{sec:shepherding}

As a large body migrates inward, it captures interior objects onto
mean motion resonances, and shepherds them inward. In the original
scenario, a gas giant forms near the snow line, and subsequently
migrates inward. As the giant migrates it shepherds interior
protoplanets inward, which collide and merge to form super Earth-mass
planets \citep{2005ApJ...631L..85Z}. Here we use $n$-body simulations
to study a similar scenario, but with a low-mass planet migrating
inwards from the snow line due to type I migration.

To investigate shepherding effects, we used the MERCURY integrator
\citep{1999MNRAS.304..793C}, including type I migration and damping
forces \citep{2006A&A...450..833C,2007A&A...472.1003F}. The migration
rate, and eccentricity and inclination damping are reduced by a factor
of $f_{\rm mig} = 10$. The inner disk edge is placed at 0.05\,AU, and
inside this point planets cease to interact with the disk
\citep{2007ApJ...654.1110T}. Simulations are initialised with a number
of isolated protoplanets in a disk between 0.1\,AU and the snow line
at $2.7\,M_\star$\,AU.

Isolation masses are calculated from Equation (\ref{eq:miso}) with the
half-spacing $B$ randomly varied between 3.75 and 4.25. We assume
Solar metallicity and $f_{\rm ice} = 10$. One protoplanet begins
beyond the snow line. This outermost protoplanet is $\approx$30 times
more massive than the one immediately interior to it
(eq. \ref{eq:miso1}). Initial eccentricities (inclinations) are
randomly distributed between 0 and 0.02 (0.5$^\circ$), and the
remaining orbital elements are randomly distributed. We set the mass
of the outermost planet at the middle of the range shown for $M_\star
= 0.25$, 0.5, and 1\,$M_\odot$ in Figure \ref{fig:f10sum}; 2, 3.2, and
6\,$M_\oplus$ respectively. Simulations are run for 10$^8$\,yr with
$\sim$0.3 day timesteps. Objects are allowed to collide, and are
assumed to merge into a single body with no fragmentation. These
simulations do not include relativistic effects, or tidal interaction
with the star. See \citet{2007ApJ...654.1110T} for a more detailed
study of migration to small radii, and how these effects affect final
system dynamics.

\subsubsection{Shepherding results}

\begin{figure}
\begin{center}
  \plotone{f7.eps}
%
%
  \caption{Migration of protoplanets in a cold finger disk for 0.25,
    0.5, and 1\,$M_\odot$ (left to right). Initial outer planet masses
    are 2, 3.2, and 6\,$M_\oplus$. Lines end in a $\circ$ when a
    collision occurs. Surviving planets more massive than
    1\,$M_\oplus$ are labelled by their final masses (in
    $M_\oplus$). Planets migrate earlier for lower stellar masses,
    because the snow line distance is closer. Objects scattered
    outward subsequently migrate, and some resume chaotic
    growth. Shepherded objects merge as their orbits are pushed
    together by the larger migrating protoplanet.}\label{fig:shep}
\end{center}
\end{figure}

Figure \ref{fig:shep} shows the semi-major axis evolution resulting
from these simulations. All show similar characteristics. Starting
from the inner disk edge, a wave of chaotic growth moves outward
\citep[e.g.][]{2001Icar..152..205C,2006AJ....131.1837K}, until the
number of protoplanets is reduced such that their spacing is
stable. This stability is set by a balance between mutual
perturbations between protoplanets, and damping by interaction with
the gas disk.

When the outermost large protoplanet begins to migrate, it scatters
the first objects it encounters into exterior orbits. When the
interaction occurs, these objects are still undergoing eccentric
chaotic growth, and are less likely to be captured onto resonances and
shepherded inward. Once scattered, the outer objects slowly migrate
inward. For 0.25 and 0.5\,$M_\odot$ the scattered planets are still
relatively close to the star, and have time to set up chains of
(mostly first order) resonant orbits. A few collisions occur. For
1\,$M_\odot$ more objects are scattered outward, which continue
chaotic growth. Despite the initial disruption by the migrating
object, $\sim$Earth-mass planets still form at $\sim$1\,AU.

When the migrating protoplanet encounters objects that have reached
stable orbits, it shepherds them inward. These smaller objects accrete
others as their orbits are pushed together, and several
$\sim$1\,$M_\oplus$ rocky objects form. Shepherded objects may be
accreted by the large migrator, or remain in interior resonant orbits
\citep[see][]{2007ApJ...654.1110T}.

These simulations show that as with the gas giant case
\citep{2005ApJ...631L..85Z}, the effect of super-Earth migration on
interior objects has observational consequences. Planets near the
outer edge of the terrestrial region are scattered outward, while
those in the inner region are shepherded to smaller radii. Shepherding
results in multiple short-period planets with different compositions.

\section{Discussion and Summary}\label{sec:summary}

We have considered two scenarios for forming short-period
$\lesssim$10\,$M_\oplus$ planets over a range of stellar masses:
planet-planet scattering and type I migration.

Our models form planets in cold finger disks. These disks have large
snow line enhancements compared to the MMSN model
\citep{1988Icar...75..146S,2004ApJ...614..490C}. Water vapour from the
terrestrial region condenses into ices outside the snow line as the
gas disk diffuses and advects. The enhancement is increased by new
water vapour delivered inside the snow line by drifting icy
planetesimals \citep{2004ApJ...614..490C}. Protoplanets forming in the
cold finger regions near the snow line are much larger than others
elsewhere in the disk.

We test the effectiveness of planet-planet scattering with $n$-body
simulations. We consider stars with masses $M_\star =
0.25$--2\,$M_\odot$ and 10\,$M_\oplus$ planets. Planets with orbits
near the limits of stability are evolved until a collision or ejection
occurs, or 1\,Gyr. Although equal mass planet-planet scattering can
produce planets with small periastra for the lowest mass stars
(Fig. \ref{fig:rfhist}), long circularisation times prevent them from
achieving circular orbits on reasonable timescales. Thus, scattering
is probably not a viable scenario for placing low-mass planets on
short-period orbits for any stellar mass. For 0.25\,$M_\odot$, planets
have periastra $\sim$0.05\,AU and semi-major axes
$\sim$0.5\,AU. Though transit durations are still several hours,
orbital periods of several hundred days and maximum radial velocities
of a few m/s make these planets hard to detect.

Migration of icy protoplanets from the snow line is a viable mechanism
for forming short-period super-Earths. Planet masses set whether they
migrate to the inner disk edge before the gas disk disperses. Some
planets with insufficient masses are stranded at intermediate radii as
the gas disk disperses; a way to form ``ocean planets''
\citep{2003ApJ...596L.105K,2004Icar..169..499L}. The minimum
protoplanet mass for migration to a close-in orbit increases as the
snow line moves out with increasing stellar mass
(Fig. \ref{fig:f10sum}). The maximum planet mass is
$\sim$10\,$M_\oplus$, because above this mass they instead accrete
large atmospheres and form gas giants.

Above $\sim$\,1\,$M_\odot$, the only protoplanets massive enough to
migrate to close-in orbits are $\gtrsim$10\,$M_\oplus$ and no hot
super-Earths form. This maximum stellar mass is independent of the
disk mass distribution, and probes type I migration efficiency. Other
uncertain parameters, such as snow line distance and disk profile do
not have major observable consequences, but are not easily constrained
by observations either.

For disks with similar snow line enhancements, the theory yields
trends with metallicity (Fig. \ref{fig:prob}). For disk masses
distributed as a power law, the frequency of short-period planets
increases with metallicity, because most disks have low
masses. However, if disk masses are distributed around a relatively
high mass, planet frequency decreases with increasing metallicity,
because planets forming in the most common disks are pushed above the
gas accretion mass at high metallicities. As planetesimal size
increases, growth slows, and becomes longer than the migration
timescale. Simulations of concurrent accretion and migration with
increased planetesimal sizes require much higher disk masses to yield
similar results.

As icy planets migrate from the snow line, they interact dynamically
with interior rocky protoplanets (Fig, \ref{fig:shep}). Protoplanets
undergoing chaotic growth are scattered onto exterior orbits. Closer
protoplanets on stable orbits damped by disk interaction are
shepherded inward, and coalesce into a few rocky objects with masses
$\sim$1\,$M_\oplus$. These objects may be accreted by the large
migrating planet, or remain as separate planets on interior
orbits. These orbits are likely near-commensurate with the icy
migrators orbit \citep{2007ApJ...654.1110T}. If planetary systems in
such configurations are found in transit surveys, compositional models
may discern differences, thus confirming their origins in rocky or icy
regions \citep{2007ApJ...665.1413V}. However, different structural
models may be degenerate if the planets have atmospheres
\citep{2007arXiv0710.4941A}.

Some planets may accrete hydrogen atmospheres due to a decreased
planetesimal accretion rate following isolation
\citep{2000ApJ...537.1013I,2006ApJ...648..666R}. To be observed as hot
super-Earths requires subsequent photoevaporation
\citep[e.g.][]{2005A&A...436L..47B}. Significant photoevaporation of
planets with massive atmospheres is unlikely unless the planet mass is
in the $\lesssim$70\,$M_\oplus$ type I migration regime
\citep{2007arXiv0711.2015R}. Thus, planets with remnant hydrogen
atmospheres may form by the same migration mechanism we present
here. Scattering is also a possibility for these planets to reach
short-period orbits, because they have higher initial masses.

For planets originating in icy regions, their largely volatile
composition has important implications for their evolution during and
after formation. Icy grains may enhance growth if they stick together
more easily, but also allows the possibility of large evaporation
events in high energy collisions of larger objects. During the violent
accretion process, and with the possible outcome of short-period
orbits, melting and evaporation of ices will affect these planets
\citep[e.g.][]{1982Icar...52...14L,2003ApJ...596L.105K,2007Icar..191..453S}.

After migrating to close-in orbits, initially icy/watery planets may
retain large super-critical steam atmospheres, or become rocky cores
stripped of volatiles entirely. \citet{2003ApJ...596L.105K} considered
the existence of volatile-rich planets in the Solar habitable zone,
and suggested that planets around Solar luminosity stars would be safe
from evaporation at $\gtrsim$1\,AU but not at closer distances. He
also noted that lower EUV luminosities for M dwarfs makes these stars
less likely to evaporate planetary atmospheres. More recently,
\citet{2007Icar..191..453S} revisited the issue, and concluded that
planets $\gtrsim$6\,$M_\oplus$ will retain most of their water content
at $\gtrsim$0.04\,AU from a Solar-type star. The results of both
studies suggest the evaporation timescale is strongly dependent on
semi-major axis. Therefore, a trend may be noticeable within the small
semi-major axis range of transiting planets.

The picture that emerges is of systems with evaporated rocky planets
inside $\sim$0.04\,AU, and steam planets somewhat outside this
distance. A few stalled ocean
\citep{2003ApJ...596L.105K,2004Icar..169..499L} and icy planets extend
through and past the habitable zone. For these planets, microlensing
provides sensitivity complementary to transit and radial velocity
methods at $\sim$AU distances \citep[e.g.][]{2006Natur.439..437B},
which will help yield trends with semi-major axis, particularly for
low-mass stars.

Surveys such as the MEarth Project \citep{2007arXiv0709.2879N}, CoRoT
\citep{2003AdSpR..31..345B}, and Kepler \citep{2003SPIE.4854..129B}
hope to discover super-Earths by the transit method. Like those
discovered by radial velocity, most planets will orbit close to their
parent stars. Because they are unlikely to form \emph{in situ}, these
planets necessarily require some form of migration or scattering from
their formation regions. Observed systems will thus test and inform
mechanisms that form and bring planets to visible orbits.

\acknowledgements

We acknowledge support from an Australian Postgraduate Award, a
Smithsonian Astrophysical Observatory pre-doctoral fellowship (GK),
and the {\it NASA Astrophysics Theory Program} through grant
NAG5-13278 and the \emph{TPF Foundation Science Program} though grant
NNG06GH25G (SK). We thank the anonymous referee for a prompt report,
which improved the content of the paper. $N$-body simulations were run
on computers maintained by the RSAA Computer Section at Mt Stromlo
Observatory.

\end{document}